% ------------------------------------------------------------------- 
%Calculation of integrals containing confluent hypergeometric functions
% -------------------------------------------------------------------
% paper Saad
% 
% integrals.tex -- in.tex [15 March 2003] [3 June 2003] 
%
% -------------------------------------------------------------------
\def\ptitle{\tiny Integrals containing confluent hypergeometric functions $\dots$}
% -------------------------------------------------------------------
%  generic unix 12 fonts (lower case names) with no magstep
% --------------------------------------------------------------------
\font\tr=cmr12                          % Our default
\font\bf=cmbx12                         % Redefinition
                         % Redefinition
\font\it=cmti12                         % Redefinition
\font\trbig=cmbx12 scaled 1500          % Main Title
                          % Theorems                       
\font\tiny=cmr10                        % Running title
% --------------------------------------------------------------------
\output={\shipout\vbox{\makeheadline
                                      \ifnum\the\pageno>1 {\hrule}  \fi 
                                      {\pagebody}   
                                      \makefootline}
                   \advancepageno}

\headline{\noindent {\ifnum\the\pageno>1 
                                   {\tiny \ptitle\hfil
page~\the\pageno}\fi}}
\footline{}
% ---------------------------------------------------------------------

\tr 
%--------------------------------------------------------------------
    % bra ket:  math mode (to replace angle)
    %   ket  >
  % new line after displayed equations
\def\ni{\noindent}             % noindent

\def\htab#1#2{{\hskip #1 in #2}}
 % bra < math mode
 % ket > math mode
\def\hi#1#2{$#1$\kern -2pt-#2} % hyphen \hi{N}{body} = N-body
\def\hy#1#2{#1-\kern -2pt$#2$} % hyphen hy{large}{N} = large-N
\def\sgn{{\rm sgn}}
%--------------------------------------------------------------------
\def\dbox#1{\hbox{\vrule % Open box size 2#1 (Abrahams p 273) 
\vbox{\hrule \vskip #1\hbox{\hskip #1\vbox{\hsize=#1}\hskip #1}\vskip #1 
\hrule}\vrule}} 
\def\qed{\hfill \dbox{0.05true in}} % QED 
 % SQUARE 

%--------------------------------------------------------------------
% SPACING
% -------------------------------------------------------------------
\baselineskip 15 true pt  % draft 15 
\parskip=0pt plus 5pt 
\parindent 0.25in
\hsize 6.0 true in 
\hoffset 0.25 true in 
% 6 in width with 1.25 in margins default = (6.5, 0)
\emergencystretch=0.6 in                 % TEXBook p 107 : allows h-space 
\vfuzz 0.4 in                            % page-length flexibility
\hfuzz  0.4 in                           % line-length flexibility
\vglue 0.1true in
\mathsurround=2pt                        % Default is 2pt
\topskip=24pt                            % Default is 10pt
% ---------------------------------------------------------------------
%  References
% ---------------------------------------------------------------------
\newcount\zz  \zz=0  % switch for printing references
\newcount\q   %  reference number
\newcount\qq    \qq=0  % starting reference number-1   (usually zero)

\def\pref#1#2#3#4#5{\frenchspacing \global \advance \q by 1     % paper reference
    \edef#1{[\the\q]}{\ifnum \zz=1{\item{$^{\the\q}$}{#2}{\bf #3},{ #4.}{~#5}\medskip} \fi}}

\def\bref #1#2#3#4#5{\frenchspacing \global \advance \q by 1     % book reference
    \edef#1{[\the\q]}
    {\ifnum \zz=1 { %
       \item{$^{\the\q}$} 
       {#2}, {\it #3} {(#4).}{~#5}\medskip} \fi}}

\def\gref #1#2{\frenchspacing \global \advance \q by 1  % general reference
    \edef#1{[\the\q]}
    {\ifnum \zz=1 { %
       \item{$^{\the\q}$} 
       {#2.}\medskip} \fi}}

 \def\sref #1{#1}

\def\references#1{\zz=#1
   \parskip=2pt plus 1pt   % default is 0pt plus 1pt       
   {\ifnum \zz=1 {\noindent \bf References \medskip} \fi} \q=\qq
%--------------------------------------------------------------------
\bref{\land}{L. D. Landau and E. M. Lifshitz}{Quantum mechanics: non-relativistic theory}{Pergamon, London, 1981}{Appendix f, page 662.}
%-----------------------------------------------------------------------------
\pref{\gord}{W. Gordon, Ann. Phys. Lpz. }{2}{1031 (1929)}{}
%-------------------------------------------------------------------------
\pref{\karu}{E. Karule, J. Phys. A: Math. Gen. }{23}{1969-1971 (1990)}{}
%-------------------------------------------------------------------------
\pref{\krw}{Krishan Sud and L. E. Wright, J. Math. Phys. }{17}{1719-1721 (1979)}{}
%-------------------------------------------------------------------------
\pref{\nor}{A. Nordsieck, Phys. Rev. }{93}{785-787 (1954)}{}
%-------------------------------------------------------------------------
\bref{\zwil}{I. S. Gradshteyn and I. M. Ryzhik}{Tables of Integrals, Series, and Products}{ 6$^{th}$ Academic Press 2000}{Specifically 7.6.}
%-----------------------------------------------------------------------------
\bref{\app}{P. Appell and J. Kamp\'e de F\'erier}{Fonctions Hyperg\'eom\'etriques et Hyp\'erspheriques }{Gauthier-Villars, Paries, France, 1926}{}
%-------------------------------------------------------------------------
\bref{\slat}{L. J. Slater}{Generalized Hypergeometric Functions}{Cambridge: Cambridge University Press 1966}{Chapter 8.}

\bref{\sriv}{H. M. Srivastava and H. L. Manocha}{A Treatise on Generating Functions}{New York: Halsted/Wiley 1984}{}

\bref{\srivk}{H. M. Srivastava and P. W. Karlsson}{Multiple Gaussian Hypergeometric Series}{New York: Halsted/Wiley 1985}{}

\pref{\erd}{A. Erd\'elyi, Math. Z. }{42}{125 (1936)}{}

\bref{\bat}{A. Erd\'elyi, W. Magnus, F. Oberhettinger, and F. G. Tricomi}{Higher Transcedental Functions}{McGraw-Hill, New York 1953}{Vol. I, Page 278.}
%-----------------------------------------------------------------------------
\pref{\wo}{W. W. Gargaro and D. S. Onley, Phys. Rev. C }{4}{1032-1043 (1971)}{}
\pref{\hah}{G.E. Hahne, J. Math. Phys. }{10}{524-531 (1969)}{}

\pref{\nas}{N. Saad and R. L. Hall, J. Phys. A: Math. Gen. }{35}{4105-4123 (2002)}{}
\bref{\slata}{L. J. Slater}{Generalized Hypergeometric Functions}{Cambridge: Cambridge University Press 1966}{Page 214, formula (8.2.3).}
%-----------------------------------------------------------------------------
\bref{\grr}{G.E. Andrews, R. Askey and R. Roy}{Special Functions}{Cambridge: Cambridge University Press 1999}{Page 65, theorem 2.2.1.}
%-----------------------------------------------------------------------------
\bref{\grrb}{G.E. Andrews, R. Askey and R. Roy}{Special Functions}{Cambridge: Cambridge University Press 1999}{Page 68, theorem 2.2.5.}
%-----------------------------------------------------------------------------
\bref{\slatb}{L. J. Slater}{Generalized Hypergeometric Functions}{Cambridge: Cambridge University Press 1966}{Page 215, formula (8.2.5).}
%-----------------------------------------------------------------------------
\bref{\slatc}{L. J. Slater}{Generalized Hypergeometric Functions}{Cambridge: Cambridge University Press 1966}{Page 218, formula (8.3.5).}
%--------------------------------------------------------------------
\bref{\landb}{L. D. Landau and E. M. Lifshitz}{Quantum mechanics: non-relativistic theory}{Pergamon, London, 1981}{formula f.10, page 664.}
%-----------------------------------------------------------------------------
\pref{\cuyt}{A. Cuyt, K. Driver, J. Tan, and B. Vertonk, J. Comp. Appl. math. }{5}{213-219 (1999)}{}
%-----------------------------------------------------------------------------
\bref{\boch}{H. Buchholz}{The Confluent Hypergeometric Function}{Springer, 1969}{Formula (4b), Page 119}
%-----------------------------------------------------------------------------
\bref{\bochb}{H. Buchholz}{The Confluent Hypergeometric Function}{Springer, 1969}{Formula (4$\beta$), Page 119}
%-----------------------------------------------------------------------------
\bref{\grrc}{G.E. Andrews, R. Askey and R. Roy}{Special Functions}{Cambridge: Cambridge University Press 1999}{Page 67, corollary 2.2.3.}
%-----------------------------------------------------------------------------
\gref{\gla}{M. L. Glasser and E. Montaldi, e-print arXiv:math.CA/9307213V1}
%-----------------------------------------------------------------------------
\pref{\hald}{R. Hall, N. Saad and A. von Keviczky, J. Phys. A: Math. Gen. }{34}{1169-1179 (2001)}{}
%-----------------------------------------------------------------------------
\pref{\hala}{R. Hall, N. Saad and A. von Keviczky, J. Math. Phys. }{39}{6345-6351 (1998)}{}
%-----------------------------------------------------------------------------
\pref{\halb}{R. Hall and N. Saad, J. Phys. A: Math. Gen. }{33}{569-578 (2000)}{}
%-----------------------------------------------------------------------------
\pref{\halc}{R. Hall and N. Saad, J. Phys. A: Math. Gen. }{33}{5531-5537 (2000)}{}
%-----------------------------------------------------------------------------
\pref{\fran}{Francisco M. Fern\'andez, Phys. Rev. A }{45}{1333-1338 (1992)}{}
%--------------------------------------------------------------------
\bref{\landf}{L. D. Landau and E. M. Lifshitz}{Quantum Mechanics: Non-Relativistic Theory}{Pergamon, London, 1981}{Problem 3, page 127.}
%--------------------------------------------------------------------
\bref{\wfr}{W. Magnus, F. Oberhettinger, and R. P. Soni}{Formulas and Theorems for the Special Functions of Mathematical Physics}{Spring-Verlag, New York 1966}{Chapter II.}

}% end of ref list

\references{0}    % Initialization of reference numbers
% ------------------------------------------------------------------ end our ref.tex

% ----------------------------
% Preprint list
% ----------------------------
\htab{3.5}{CUQM-98}

\htab{3.5}{math-ph/0306043}

\htab{3.5}{June 2003}
%-------------------------------------------------------------------
%Title Page 
%-------------------------------------------------------------------
%-------------------------------------------------------------------
%Title Page 
%-------------------------------------------------------------------
\vskip 0.5 true in
\centerline{\bf\trbig Integrals containing confluent hypergeometric functions}
\vskip 0.2 true in
\centerline{\bf\trbig  with applications to perturbed singular potentials}
\medskip
\vskip 0.25 true in
\centerline{Nasser Saad$^\dagger$ and Richard L. Hall$^\ddagger$}
\bigskip
{\leftskip=0pt plus 1fil
\rightskip=0pt plus 1fil\parfillskip=0pt
\obeylines
$^\dagger$Department of Mathematics and Statistics,
University of Prince Edward Island, 
550 University Avenue, Charlottetown, 
PEI, Canada C1A 4P3.\par}

\medskip
{\leftskip=0pt plus 1fil
\rightskip=0pt plus 1fil\parfillskip=0pt
\obeylines
$^\ddagger$Department of Mathematics and Statistics, Concordia University,
1455 de Maisonneuve Boulevard West, Montr\'eal, 
Qu\'ebec, Canada H3G 1M8.\par}

\vskip 0.5 true in
%---------------------------------------------------------------------------
% Abstract
%---------------------------------------------------------------------------
\centerline{\bf Abstract}\medskip
We show that many integrals containing products of confluent hypergeometric functions follow directly from one single integral that has a very simple formula in terms of Appell's double series $F_2$.  We present some techniques for computing such series.  Applications requiring the matrix elements of singular potentials and the perturbed Kratzer potential are presented.
\vfil\eject

%---------------------------------------------------------------------------
% 1. Introduction
%---------------------------------------------------------------------------
\ni{\bf 1. Introduction}
\medskip
\noindent Landau and Lifshitz \sref{\land} have discussed an important type of integral containing a pair of confluent hypergeometric functions ${}_1F_1,$ namely
$$J_\gamma^{sp}(a,a^\prime)=\int\limits_0^\infty t^{\gamma-1+s}e^{-h t} {}_1F_1(a;\gamma;k t){}_1F_1(a^\prime;\gamma-p;k^\prime t) dt,\eqno(1.1)
$$
where $s$ and $p$ are positive integers, and values of the parameters are supposed such that the integral converges absolutely; $Re(h)>0$. These authors mentioned that a general formula for such integrals can be derived by means of a method proposed earlier by Gordon \sref{\gord}, but this is so complex that it cannot be conveniently used.  Instead, they investigate the particular case $J_\gamma^{00}(\alpha,\alpha^\prime)$ and they show that, for $h={k+k^\prime\over 2}$,
$$J_\gamma^{00}(a,a^\prime)= 2^\gamma\Gamma(\gamma)(k+k^\prime)^{a+a^\prime-\gamma}(k^\prime-k)^{-a}(k-k^\prime)^{-a^\prime}{}_2F_1(a,a^\prime;\gamma;-{4kk^\prime\over (k^\prime-k)^2}).\eqno(1.2)
$$
The functions  ${}_1F_{1}$ and ${}_2F_1$ just mentioned are particular cases of the generalized hypergeometric function 
$$
{}_pF_{q}(\alpha_1,\alpha_2,\dots,\alpha_p;\beta_1,\beta_2,\dots,\beta_q;x)=\sum\limits_{k=0}^\infty 
{\prod\limits_{i=1}^p(\alpha_i)_k\over  \prod\limits_{j=1}^q(\beta_j)_k}{x^k\over k!},\eqno(1.3)
$$ 
where $p$ and $q$ are non-negative integers, and none of the parameters $\beta_j$ ($j=1,2,\dots,q$) is equal to zero or to a negative integer. It is known that if the series does not terminate (that is to say, if none of the $\alpha_i$, $i=1,2,\dots,p$, is a negative integer), then, in the case $p=q+1$, the series converges or diverges according as $|x|<1$ or $|x|>1$. For $x=1$ on the other hand, the series is convergent, provided
$
{\sum\limits_{j=1}^q \beta_j-\sum\limits_{i=1}^p \alpha_i}>0.
$ 
Note $(\delta)_n$, the shifted factorial (or {\it Pochhammer symbol}), is defined by
$$(\delta)_m={\Gamma(\delta+m)\over \Gamma(\delta)}=\cases{1,&if $m=0$\cr
	\delta(\delta+1)\dots (\delta+m-1),&if $m=1,2,\dots$\cr
	}\eqno(1.4)
$$

\noindent Karule \sref{\karu}, in order to overcome the inconvenient approach suggested by Landau and Lifshitz, proposed a method of derivation which allowed him to obtain a general formula for the integrals (1.1). He showed that
$$\eqalign{J_\gamma^{sp}(\alpha,\alpha^\prime)&=(1-{k\over h})^{-\alpha}(1-{k^\prime\over h})^{-\alpha^\prime}\sum\limits_{m=0}^{s+p}\bigg[{(-s-p)_m(\alpha^\prime)_m\over (c-p)_m~m!}\bigg(1-{h\over k^\prime}\bigg)^{-m}\times\cr
&\sum\limits_{r=0}^{s+m}{(-s-m)_r(\alpha)_r \over(c)_r~r!}\bigg(1-{h\over k}\bigg)^{-r} {}_2F_1(\alpha+r,\alpha^\prime+m;c+r;{kk^\prime\over (h-k^\prime)(h-k)})\bigg].} \eqno(1.5)$$
It should be clear however that the integral (1.1) and consequently (1.5) is convergent for $|k|+|k^\prime|<|h|$. For $h= {k+k^\prime\over 2}$, both (1.2) and (1.5) should be interpreted in terms of analytic continuations whenever the integral (1.1) is absolutely converge. For example, if $a$ and $a^\prime$ are nonpositive integers and since the confluent hypergeometric functions ${}_1F_1$ becomes polynomials, the condition $|h|> |k|+|k^\prime|$ is no longer necessary and the condition $Re(h)>0$ is sufficient for the convergent of the integral to take a place. 

In the present article, we show that the integral (1.1) is indeed a special case of a more general type which arises in the different fields in physics. For example, the computation of the radial matrix elements of the electromagnetic interaction between the states of a relativistic electron in the Coulomb field of a point nucleus \sref{\krw}, namely
$$
\int\limits_0^\infty t^{d-1}e^{-ht} {}_1F_1(a;b;k t){}_1F_1(a^\prime;b^\prime;k^\prime t)~dt.\eqno(1.6)
$$
The matrix element integral \sref{\nor} for bremsstrahlung or pair production without Born approximation is usually expressed in terms of confluent hypergeometric functions (1.6).

For convenience, in section 2, we present an elementary method to find a general formula for the integral (1.6) in terms of the Appell's series $F_2(d;a,a^\prime;b,b^\prime;{k\over h},{k^\prime\over h})$. The series that is absolutely convergent if $|k|+|k^\prime|<|h|$. Furthermore, we show that many integrals \sref{\zwil} involving confluent hypergeometric functions, or a product of confluent hypergeometric functions with weight measure $\{t^{d-1}e^{-ht}\},$ follow immediately from the integral (1.6). In section 3, we prove that the result of Karule, (1.5), follows for special values of the parameters in the Appell's series $F_2$. This allows us to develop some new results useful for the computation of Appell's series $F_2$. For example, we shall give a generalization of the known identity
$$F_2(d;a,a^\prime;d,d;x,y)=(1-x)^{-a}(1-y)^{-a^\prime}{}_2F_1(a,a^\prime;d;{xy\over (1-x)(1-y)})$$
in terms of the Gauss hypergeometric series ${}_2F_1$. We also present some techniques for computing the Appell's hypergeometric series $F_2$, whereby we discuss several recurrence relations for $F_2$. In section 4, we present some applications involving the computation of the matrix elements for singular potentials and for the perturbed Kratzer potential.
\medskip
%---------------------------------------------------------------------------
% 2. General type of integrals
%---------------------------------------------------------------------------
\ni{\bf 2. An integral of general type and some special cases}
\medskip
\noindent Appell's hypergeometric \sref{\app-\srivk} series $F_2$, which is a function of two complex arguments $x$, and $y$, and of five complex parameters $\alpha, a_1,a_2,b_1,$ and $b_2$ ($b_1,b_2\neq 0,-1,-2,\dots$), is defined by  
$$F_2(\alpha;a_1,a_2;b_1,b_2;x,y)=\sum\limits_{m=0}^\infty\sum\limits_{n=0}^\infty {(\alpha)_{m+n}(a_1)_m(a_2)_n\over (b_1)_m(b_2)_n~m!~n!}x^m y^n.\eqno(2.1)$$
where the Pochhammer symbols $(\delta)_n$ are defined by (1.4). The series converges absolutely for $|x|+|y|<1$ and, in general, diverges for $|x|+|y|>1$. In the present work, we will assume, unless otherwise stated, that $a_1,a_2,b_1,$ and $b_2$ are real with  $b_1$ and $b_2$ strictly positive. The following lemma can be seen as an integral representation of $F_2,$ as well as a Laplace transform of the product of two confluent hypergeometric functions \sref{\erd}.
\medskip
%-----------------------
\noindent {\bf Lemma 1:}
%-----------------------

\noindent {\it
For $Re(d) > 0$, and $|k|+|k^\prime|<|h|$, 
$$\int\limits_0^\infty t^{d-1}e^{-ht} {}_1F_1(a;b;k t){}_1F_1(a^\prime;b^\prime;k^\prime t)~ dt=
h^{-d}~\Gamma(d)~F_2(d;a,a^\prime;b,b^\prime;{k\over h},{k^\prime\over h})\eqno(2.2)
$$}
\noindent{Proof:} From the series representation (1.3) of the confluent hypergeometric series ${}_1F_1$, we have
$$\eqalign{
\int\limits_0^\infty t^{d-1}e^{-ht} {}_1F_1(a;b;k t){}_1F_1(a^\prime;b^\prime;k^\prime t) dt&=
\int\limits_0^\infty t^{d-1} e^{-ht} \bigg[\sum\limits_{m=0}^\infty \sum\limits_{n=0}^\infty{(a)_m(a^\prime)_n\over (b)_m(b^\prime)_n}
{k^m{k^\prime}^n\over m! n!}t^{n+m}\bigg]dt\cr
&{\buildrel (1)\over =}
\sum\limits_{m=0}^\infty \sum\limits_{n=0}^\infty{(a)_m(a^\prime)_n\over (b)_m(b^\prime)_n}
{k^m{k^\prime}^n\over m! n!}\int\limits_0^\infty t^{d+n+m-1}e^{-ht}dt\cr
&{\buildrel (2)\over =}
 h^{-d}\Gamma(d)\sum\limits_{m=0}^\infty \sum\limits_{n=0}^\infty{(d)_{m+n}(a)_m(a^\prime)_n\over (b)_m(b^\prime)_n~m!~ n!}
\bigg({k^\prime\over h}\bigg)^n \bigg({k\over h}\bigg)^m\quad\quad(2.3) \cr
}$$
where $(1)$ is justify by the condition $|k|+|k^\prime|<|h|$ and the fact that for large $x$ the confluent hypergeometric function ${}_1F_1(a,c,x)$ is asymptotic to \sref{\bat}
$${}_1F_1(a,c,x)={\Gamma(c)\over \Gamma(a)}e^xx^{a-c}[1+O(|x|^{-1}];$$ 
$(2)$ follows by means of the definition of Gamma function with $Re(d) > 0$ and $ Re(h) > 0,$ and the use of Pochhammer's identity
$$\Gamma(d+m+n)=(d+m)_n\Gamma(d+m)= (d+m)_n(d)_m\Gamma(d).$$
 The right hand side of (2.3) is Appell's hypergeometric series $F_2$ which observation completes the proof of the lemma.\qed
\medskip
Unfortunately, the direct use of the series representation (2.1) for the Appell's hypergeometric series $F_2$ does not converge for the variables needed in the computation of Landau and Lifshitz's integral (1.1), namely $h<k+k^\prime$. However, there are several analytic continuations of Appell's hypergeometric series $F_2$ available in the literature \sref{\krw}, \sref{\bat-\hah}. In the next section we shall give several new analytic continuation of $F_2$ that can be used directly in the computation of the integral (1.1). First, we show that lemma 1 can be used directly to prove many  standard and non-standard results. The Appell's $F_2(d;a,a^\prime;c,c^\prime;x,y)$ function reduces to a Gauss hypergeometric function ${}_2F_1$ if $a$ or $a^\prime$ or just one of the variables $(x,y)$ is zero.
\medskip
%-----------------------
\noindent {\bf Lemma 2:}
%-----------------------
\item{1.} $F_2(d;0,a^\prime;c,c^\prime;x,y)={}_2F_1(d,a^\prime;c^\prime,y)$.
\item{2.} $F_2(d;a,0;c,c^\prime;x,y)={}_2F_1(d,a;c,x)$.
\item{3.} $F_2(d;a,a^\prime;c,c^\prime;0,y)={}_2F_1(d,a^\prime;c^\prime,y)$.
\item{4.} $F_2(d;a,a^\prime;c,c^\prime;x,0)={}_2F_1(d,a;c,x)$.
\medskip
\noindent These allow us to deduce the following known formula, for $Re(d)>0$ and $|k|<|h|$,
$$\int\limits_0^\infty t^{d-1}e^{-ht} {}_1F_1(a;b;k t)~ dt = h^{-d}~\Gamma(d)~{}_2F_1(d,a;b;{k\over h}).\eqno(2.4)$$
Many identities follow directly from (2.4) by means of the expansion for ${}_2F_1$ in terms of elementary functions \sref{\bat}, a few of which we collect in Appendix I for use in further applications.  Many other formulae can be deduced directly from (2.2), for example, formulas (19) to (21) in Appendix I.
\medskip
%-----------------------
\noindent {\bf Lemma 3:}
%-----------------------
\item{1.} $F_2(d;a,a^\prime;a,a^\prime;x,y)=(1-x-y)^{-d}$
\item{2.} $F_2(d;a,a^\prime;d,a^\prime;x,y)=(1-y)^{a-d}(1-x-y)^{-a}$
\item{3.} $F_2(d;a,a^\prime;a,d;x,y)=(1-x)^{a^\prime-d}(1-x-y)^{-a^\prime}$
\medskip
\noindent Consequently, the following formulae follow:
for $Re(d)>0$ and $|k|+|k^\prime|<|h|$, we have
$$
\int\limits_0^\infty t^{d-1}e^{-ht} {}_1F_1(a;a;k t){}_1F_1(a^\prime;a^\prime;k^\prime t)~ dt=
\Gamma(d)(h-k-k^\prime)^{-d}\eqno(2.5)
$$
which is expected since ${}_1F_1(\beta;\beta;x)=e^x$. Also, for $Re(d)>0$ and $|k|+|k^\prime|<|h|$, we have
$$
\int\limits_0^\infty t^{d-1}e^{-ht} {}_1F_1(a;d;k t){}_1F_1(a^\prime;a^\prime;k^\prime t)~dt=
\Gamma(d)(h-k^\prime)^{a-d}(h-k-k^\prime)^{-a}.\eqno(2.6)
$$

Recently, Cuyt et al \sref{\cuyt} developed a finite sum representation of Appell's hypergeometric series \sref{\app-\srivk}
$$F_1(a;b,b^\prime;c;x,y)=\sum\limits_{m=0}^\infty\sum\limits_{n=0}^\infty {(\alpha)_{m+n}(b)_m(b^\prime)_n\over (c)_{m+n}~m!~n!} x^m y^n,\eqno(2.7)$$
under the conditions that the parameters $a,b,b^\prime,$ and $c$ are positive integers; namely 
\medskip
%-----------------------
\noindent {\bf Lemma 4:}
%-----------------------
{\it For any non-negative integers $a,s,t,d$ and $x\neq y$, we have for $|x|<1$, $|y|<1$,
$$\eqalign{{\Gamma(a+1)\Gamma(d+1)\over \Gamma(a+d+2)} &F_1(a+1,s+1,t+1;a+d+2;x,y)=\cr
&\sum_{m=0}^d \pmatrix{d \cr m \cr}(-1)^m\bigg\{y^{s-a-m}\sum_{j=0}^t \pmatrix{j+s\cr s \cr} {(-x)^j\over (y-x)^{s+j+1}}A\cr
&+x^{t-a-m}\sum_{k=0}^s \pmatrix{k+t\cr t \cr} {(-y)^k\over (x-y)^{t+k+1}}C\bigg\}}\eqno(2.8)
$$
where
$$A=\sum_{k=0,k\neq t-j}^{a+m} \pmatrix{a+m\cr k \cr}(-1)^k{[1-(1-y)^{k+j-t}]\over (k+j-t)}-\pmatrix{a+m\cr t-j \cr}(-1)^{t-j}\ln(1-y),$$
$$C=\sum_{j=0,j\neq s-k}^{a+m} \pmatrix{a+m\cr j \cr}(-1)^j{[1-(1-x)^{j+k-s}]\over (j+k-s)}-\pmatrix{a+m\cr s-k \cr}(-1)^{s-k}\ln(1-x).$$
Here, $\pmatrix{m\cr k \cr}$ is the binomial coefficient and all sums are understood to be zero where they are not defined.}
\medskip
These results can be used along with the following reduction formulae for $F_2$ to obtain many new identities for the integral (2.1) with positive integer parameters. \medskip
%-----------------------
\noindent {\bf Lemma 5:}
%-----------------------
\item{1.} $F_2(d;a,a^\prime;d,c^\prime;x,y)=(1-x)^{-a}F_1(a^\prime,a,d-a;c^\prime;{y\over 1-x},y)$
\item{2.} $F_2(d;a,a^\prime;c,d;x,y)=(1-y)^{-a^\prime}F_1(a,d-a^\prime,a^\prime;c;x,{x\over 1-y})$
\medskip
\noindent For example; since
$$
\eqalign{F_2(4;2,1;4,2;x,y)&=(1-x)^{-2}F_1(1,2,2;2;{y\over 1-x},y)\cr
&={1\over x^2(1-y)}+{1\over x^2(1-x)(1-x-y)}-{2\over x^3y}\bigg[\ln(1-y)-\ln(1-{y\over 1-x})\bigg]}\eqno(2.8)$$
we have
$$\eqalign{\int\limits_0^\infty t^{3}e^{-ht}{}_1F_1(2;4;kt){}_1F_1(1;2;&k^\prime t)dt=\Gamma(4)\bigg\{{1\over hk^2(h-k^\prime)}+{1\over k^2(h-k)(h-k-k^\prime)}\cr
&-{2\over k^3k^\prime}\bigg[\ln(h-k^\prime)+\ln(h-k)-\ln(h)-\ln(h-k-k^\prime)\bigg]\bigg\}}\eqno(2.9)
$$
\medskip
%---------------------------------------------------------------------------
% 3. Calculations of $F_2$
%---------------------------------------------------------------------------
\ni{\bf 3. Calculations of Appell's hypergeometric function $F_2$}
\medskip

In this section, we present our main results concerning the computation of $F_2$ and its analytic continuation.  The integral formulae in this section are mostly new, as distinct from the many known results which we can recover, some of which we have listed in the appendix.
\medskip
%-----------------------
\noindent {\bf Lemma 6:}
%-----------------------
{\it For $s$ and $d$ positive integers, 
\item{1.} In terms of $F_1$, we have
$$
\eqalign{F_2(c+s;a,a^\prime;c,c-p;{k\over h},{k^\prime\over h})&=
\bigg(1-{k^\prime\over h}\bigg)^{-a^\prime}\sum\limits_{m=0}^{s+p}{(a^\prime)_m(-s-p)_m\over (c-p)_m~m!}
\bigg(1-{h\over k^\prime}\bigg)^{-m}\cr
&\times F_1(a;c+s-a^\prime,m+a^\prime;c;{k\over h};{k\over h-k^\prime}).}
\eqno(3.1)$$
\item{2.} In terms of ${}_2F_1$, we have
$$\eqalign{F_2(c+s;&a,a^\prime;c,c-p;{k\over h},{k^\prime\over h})=\bigg(1-{k\over h}\bigg)^{-a}\bigg(1-{k^\prime\over h}\bigg)^{-a^\prime}\sum\limits_{m=0}^{s+p}\bigg[{(a^\prime)_m(-s-p)_m\over (c-p)_m~m!}
\bigg(1-{h\over k^\prime}\bigg)^{-m}\cr
&\times\sum\limits_{r=0}^{s+m}{(a)_r(-s-m)_r~\over (c)_r~r!}\bigg(1-{h\over k}\bigg)^{-r} {}_2F_1(a+r;a^\prime+m;c+r;{kk^\prime\over (h-k^\prime)(h-k)})\bigg]}\eqno(3.2)$$
where $|k|+|k|^\prime<|h|$.
}
\medskip
\noindent{Proof:} From the integral representation \sref{\slata} of the Appell's hypergeometric function $F_2$, we have
$$\eqalign{F_2(c+s;a,a^\prime;c,c-p;&{k\over h},{k^\prime\over h})={\Gamma(c)\Gamma(c-p)\over \Gamma(a)\Gamma(a^\prime)\Gamma(c-a)\Gamma(c-p-a^\prime)}\cr
&\int\limits_0^1\int\limits_0^1u^{a-1}v^{a^\prime-1}(1-u)^{c-a-1}(1-v)^{c-p-a^\prime-1}(1-{k\over h}u-{k^\prime\over h}v)^{-c-s}dudv}
$$
under the conditions: $c>a>0, c-p>a^\prime>0$.
We shall look first at the integral
$$\int\limits_0^1 v^{a^\prime-1}(1-v)^{c-p-a^\prime-1}(1-{ku\over h}-{k^\prime v\over h})^{-c-s}dv
=(1-{ku\over h})^{-c-s}\int\limits_0^1 v^{a^\prime-1}(1-v)^{c-p-a^\prime-1}(1-{k^\prime v\over h-ku})^{-c-s}dv.
$$
Since the integral on the right-hand side is now the integral representation \sref{\grr} of the Gauss hypergeometric function ${}_2F_1$, we have
$$\eqalign{
\int\limits_0^1 v^{a^\prime-1}(1-v)^{c-p-a^\prime-1}\bigg(1-{ku\over h}-{k^\prime v\over h}\bigg)^{-c-s}dv
&={\Gamma(a^\prime)\Gamma(c-p-a^\prime)\over \Gamma(c-p)} \bigg(1-{ku\over h}\bigg)^{-c-s}\cr
&\quad \times {}_2F_1(c+s,a^\prime;c-p;{k^\prime\over h-ku}).}
$$
From Pfaff's transformation \sref{\grrb} of ${}_2F_1$, namely
$$
{}_2F_1(a,b;c;z)=(1-z)^{-a}{}_2F_1(a,c-b;c;{z\over z-1}),\eqno(3.3)
$$
we have
$$\eqalign{{}_2F_1(c+s,a^\prime;c-p;{k^\prime\over h-ku})
&{\buildrel (1)\over =}(1-{k^\prime\over h-ku})^{-a^\prime}{}_2F_1(a^\prime,-p-s;c-p;{k^\prime\over -h+ku+k^\prime})\cr
&=(1-{k^\prime\over h-ku})^{-a^\prime}\sum\limits_{m=0}^{s+p}{(a^\prime)_m(-s-p)_m\over (c-p)_m~m!}({k^\prime\over -h+ku+k^\prime})^m
}
$$
since ${}_2F_1$ on the right-hand side of $(1)$ is a polynomial of order ${s+p}$. Therefore, we may now write
$$\eqalign{F_2(c+s;a,a^\prime;c,c-p;&{k\over h},{k^\prime\over h})={\Gamma(c)\over \Gamma(a)\Gamma(c-a)}h^{c+s}\sum\limits_{m=0}^{s+p}{(a^\prime)_m(-s-p)_m\over (c-p)_m~m!}({-k^\prime})^m
\cr
&\times\int\limits_0^1 u^{a-1}(1-u)^{c-a-1}(h-ku)^{-c-s+a^\prime}(h-ku-k^\prime)^{-m-a^\prime}du\cr
&={\Gamma(c)\over \Gamma(a)\Gamma(c-a)}({h\over h-k^\prime})^{a^\prime}\sum\limits_{m=0}^{s+p}{(a^\prime)_m(-s-p)_m\over (c-p)_m~m!}({k^\prime\over k^\prime-h})^m
\cr
&\times\int\limits_0^1 u^{a-1}(1-u)^{c-a-1}(1-{k\over h}u)^{-c-s+a^\prime}(1-{k u\over h-k^\prime})^{-m-a^\prime}du}.
$$
The last integral is an integral representation \sref{\slatb} of Appell's series $F_1$, namely
$$F_1(a;c+s-a^\prime,m+a^\prime;c;{k\over h};{k\over h-k^\prime})={ \Gamma(c)\over \Gamma(a)\Gamma(c-a)}\int\limits_0^1 u^{a-1}(1-u)^{c-a-1}(1-{k\over h}u)^{-c-s+a^\prime}(1-{k u\over h-k^\prime})^{-m-a^\prime}du
$$
Therefore
$$F_2(c+s;a,a^\prime;c,c-p;{k\over h},{k^\prime\over h})=({h\over h-k^\prime})^{a^\prime}\sum\limits_{m=0}^{s+p}{(a^\prime)_m(-s-p)_m\over (c-p)_m~m!}({k^\prime\over k^\prime-h})^m F_1(a;c+s-a^\prime,m+a^\prime;c;{k\over h};{k\over h-k^\prime}).$$
This proof (3.1) in the lemma. From the identity \sref{\slatc}
$$F_1(a;b,b^\prime;c;x,y)=(1-x)^{-a}F_1(a;c-b-b^\prime,b^\prime;c;{x\over x-1},{y-x\over 1-x})$$
we have
$$\eqalign{F_2(c+s;a,a^\prime;c,c-p;{k\over h},{k^\prime\over h})&=(1-{k\over h})^{-a}(1-{k^\prime\over h})^{-a^\prime}\sum\limits_{m=0}^{s+p}{(a^\prime)_m(-s-p)_m\over (c-p)_m~m!}
(1-{h\over k^\prime})^{-m}\times\cr
&\quad\quad F_1(a;-s-m,a^\prime+m;c;{k\over k-h};{kk^\prime\over (h-k^\prime)(h-k)})}$$
As the second parameter of the $F_1$ function is a negative integer, we may now write 
$$\eqalign{F_2(c+s;a,a^\prime;c,c-p;{k\over h},{k^\prime\over h})&=(1-{k\over h})^{-a}(1-{k^\prime\over h})^{-a^\prime}\sum\limits_{m=0}^{s+p}{(a^\prime)_m(-s-p)_m\over (c-p)_m~m!}
(1-{h\over k^\prime})^{-m}\times\cr
&\sum\limits_{r=0}^{s+m}{(a)_r(-s-m)_r~k^r\over (c)_r(k-h)^r~r!} {}_2F_1(a+r;a^\prime+m;c+r;{kk^\prime\over (h-k^\prime)(h-k)})}$$
because of the identity
$$F_1(a;b,b^\prime;c;x,y)=\sum\limits_{r=0}^\infty {(a)_r(b)_r x^r\over (c)_r~r!}{}_2F_1(a+r,b^\prime;c+r;y)
$$
which complete the proof of the lemma.\qed
\medskip
We may also apply Pfaff's transformation (3.3) to the hypergeometric function ${}_2F_1$ in (3.2) which allow us to write (3.2) as follows
$$\eqalign{
F_2(c+s;&a,a^\prime;c,c-p;{k\over h},{k^\prime\over h})=(1-{k\over h})^{-a}
(1-{k^\prime\over h})^{-a^\prime}\sum\limits_{m=0}^{s+p}\bigg[{(-s-p)_m(a^\prime)_m\over (c-p)_m~m!}(1-{h\over k^\prime})^{-m}\times\cr
&\sum\limits_{r=0}^{s+m}
{(a)_r(-s-m)_r~k^r\over (c)_r~(k-h)^r~r!}(1-{kk^\prime\over (h-k^\prime)(h-k)})^{-a^\prime-m}{}_2F_1(c-a,a^\prime+m;c+r;
{kk^\prime\over h(h-k-k^\prime)})\bigg]}$$
As an immediate result of (3.2) the following identity follows:
$$F_2(c;a,a^\prime;c,c;x,y)=(1-x)^{-a}(1-y)^{-a^\prime}{}_2F_1(a,a^\prime;c;{xy\over (1-x)(1-y)}).\eqno(3.4)$$
\medskip
A general formula for Landau and Lifshitz's integral type (1.1) follows by means of lemma 1 and lemma 6, where in this case we have 
\medskip
%-----------------------
\noindent {\bf Lemma 7:}
%-----------------------
{\it With ${k+k^\prime\over 2}=h$, 
$$J_\gamma^{sp}(\alpha,\alpha^\prime)
=
h^{-\gamma-s}\Gamma(\gamma+s)F_2(\gamma+s;a,a^\prime;\gamma,\gamma-p;{k\over h},{k^\prime\over h})\eqno(3.5)
$$
where $F_2(\gamma+s;a,a^\prime;\gamma,\gamma-p;{k\over h},{k^\prime\over h})$ is now computed by means of lemma 6 for $s$ and $p$ positive integers, and $Re(h) > 0$. Furthermore (1.2) follows directly for the special case $s=p=0$.}
\medskip
We may apply a similar proof of lemma 6 to prove the following result concerning the calculation of function $F_2$.
\medskip
%-----------------------
\noindent {\bf Lemma 8:}
%-----------------------
{\it For $s$ and $d$ positive integers and $s\geq p$, 
\item{1.} In terms of $F_1$, we have
$$\eqalign{
F_2(c+s;&a,a^\prime;c,c+p;{k\over h},{k^\prime\over h})=(1-{k\over h})^{-a}
(1-{k^\prime\over h})^{-a^\prime}\times\cr
&\sum\limits_{m=0}^{s-p}{(p-s)_m(a^\prime)_m\over (c+p)_m~m!}(1-{h\over k^\prime})^{-m}F_1(a;-s-m,m+a^\prime;c;{k\over k-h};{kk^\prime\over (h-k^\prime)(h-k)})}\eqno(3.6)$$
\item{2.} In terms of ${}_2F_1$, we have
$$\eqalign{F_2(c+s;&a,a^\prime;c,c+p;{k\over h},{k^\prime\over h})=(1-{k\over h})^{-a}(1-{k^\prime\over h})^{-a^\prime}\sum\limits_{m=0}^{s-p}{(p-s)_m(a^\prime)_m\over (c+p)_m~m!}
(1-{h\over k^\prime})^{-m}\times\cr
&\sum\limits_{r=0}^{s+m}{(a)_r(-s-m)_r~\over (c)_r~r!}(1-{h\over k})^{-r} {}_2F_1(a+r;a^\prime+m;c+r;{kk^\prime\over (h-k^\prime)(h-k)})}\eqno(3.7)$$
where $|k|+|k|^\prime<|h|$.
}\medskip

\noindent In particular, if $s=p$ we have 
$$
F_2(c+s;a,a^\prime;c,c+s;{k\over h},{k^\prime\over h})=(1-{k\over h})^{-a}
(1-{k^\prime\over h})^{-a^\prime}F_1(a;-s,a^\prime;c;{k\over k-h},{kk^\prime\over (h-k^\prime)(h-k)}).\eqno(3.8)$$
Furthermore, from lemma 8, we have
$$\eqalign{F_2(c+s;&a,a^\prime;c,c+s;{k\over h},{k^\prime\over h})=\bigg(1-{k\over h}\bigg)^{-a}\bigg(1-{k^\prime\over h}\bigg)^{-a^\prime}\cr
&\sum\limits_{r=0}^{s}{(a)_r(-s)_r~\over (c)_r~r!}\bigg(1-{h\over k}\bigg)^{-r} {}_2F_1(a+r;a^\prime;c+r;{kk^\prime\over (h-k^\prime)(h-k)})}\eqno(3.9)$$
\medskip
We can also apply Lemma 2 to obtain some new results concerning the calculations of the Appell's hypergeometric function $F_2$.
\medskip 
%-----------------------
\noindent {\bf Lemma 9:} 
%-----------------------
{\it For $|k|+|k^\prime|<|h|$, we have
$$F_2(d;a,a;c,c;{k\over h},{k^\prime\over h})=\sum\limits_{r=0}^\infty {(a)_r(c-a)_r(d)_{2r}\over (c)_r~(c)_{2r}~r!} \bigg(-{kk^\prime\over h^2}\bigg)^r{}_2F_1(d+2r,a+r;c+2r;{k+k^\prime\over h})\eqno(3.10)$$
In particular, if $k^\prime=-k$
$$
F_2(d;a,a;c,c;{k\over h},-{k\over h})={}_4F_3(a,c-a,{d\over 2},{d+1\over 2};c,{c\over 2},{c+1\over 2};{k^2\over h^2})\quad\quad \hbox{ for }\quad 2|k|<|h|.\eqno(3.11)
$$
}\medskip
\noindent{Proof:} Form the following identity for the product of confluent hypergeometric functions \sref{\gla}
$${}_1F_1(a;c;x){}_1F_1(a;c;y)=\sum\limits_{r=0}^\infty{(a)_r(c-a)_r\over (c)_r~(c)_{2r}~r!} (-xy)^r{}_1F_1(a+r;c+2r;x+y)\eqno(3.12)$$
we replace $x$ and $y$ by ${kt\over h}$ and  ${k^\prime t\over h}$ respectively, then integrate with respect to $t$ both sides after multiplying through by $t^{d-1}e^{-ht}$. The lemma follows immediately by applying (2.2) and (2.4) to the both sides.

We can  proof the particular case by re-placing $k^\prime$ with $-k$ in (3.10) and applying the Gauss's duplication formula
$$(a)_{2n}=2^{2n}\bigg({a\over 2}\bigg)_n\bigg({a+1\over 2}\bigg)_n$$
and then using the series representation expansion for ${}_4F_3$ as given by (1.3).\qed 
\medskip
\noindent The lamme can be now used to show that: for $Re(d)>0$ and $2|k|<|h|$, we have
$$
\int\limits_0^\infty t^{d-1}e^{-ht} {}_1F_1(a;b;k t){}_1F_1(a;b;-k t)~dt=
{\Gamma(d)\over h^d}~{}_4F_3({d\over 2},{d+1\over 2},a,b-a;b,{b\over 2},{b+1\over 2};{k^2\over h^2})\eqno(3.13)
$$
or equivalently
$$
\int\limits_0^\infty t^{d-1}e^{-ht} {}_2F_3(a,b-a;b,{b\over 2},{b+1\over 2};{k^2t^2\over 4} )~dt=
{\Gamma(d)\over h^d}~{}_4F_3({d\over 2},{d+1\over 2},a,b-a;b,{b\over 2},{b+1\over 2};{k^2\over h^2})\eqno(3.14)
$$
as a result of Ramanujan's theorem:
$${}_1F_1(a;b;x){}_1F_1(a;b;-x)={}_2F_3(a,b-a;b,{b\over 2},{b+1\over 2};{x^2\over 4}).\eqno(3.15)$$

The recurrence relations of the confluent hypergeometric functions can be used, by means of lemma 1, to develop several recurrence relations for the Appell's hypergeometric function $F_2$. For example, using the recurrence relation
$$(b-a){}_1F_1(a-1;b;z)+(2a-b+z){}_1F_1(a;b;z)-a{}_1F_1(a+1;b;z)=0,\eqno(3.16)$$
then, for $Re(d)>-1$, the recurrence relation for $F_2$ follows
$$\eqalign{{k\over h}(d-1)F_2(d;a,a^\prime;b,b^\prime;{k\over h},{k^\prime\over h})
&=(a-b)F_2(d-1;a-1,a^\prime;b,b^\prime;{k\over h},{k^\prime\over h})\cr&+(b-2a)F_2(d-1;a,a^\prime;b,b^\prime;{k\over h},{k^\prime\over h})\cr
&+aF_2(d-1;a+1,a^\prime;b,b^\prime;{k\over h},{k^\prime\over h}).}\eqno(3.17)$$
In particular, if $d=b^\prime+s$ where $s$ is positive integer, we have
$$\eqalign{{k\over h}(b^\prime+s-1)F_2(b^\prime+s;a,a^\prime;b,b^\prime;{k\over h},{k^\prime\over h})
&=(a-b)F_2(b^\prime+s-1;a-1,a^\prime;b,b^\prime;{k\over h},{k^\prime\over h})\cr&+(b-2a)F_2(b^\prime+s-1;a,a^\prime;b,b^\prime;{k\over h},{k^\prime\over h})\cr
&+aF_2(b^\prime+s-1;a+1,a^\prime;b,b^\prime;{k\over h},{k^\prime\over h}).}\eqno(3.18)$$
Consequently, if $s=1$, the following recurrence relation follows
$$\eqalign{{b^\prime k\over h}F_2(b^\prime+1;a,a^\prime;b,b^\prime;{k\over h},{k^\prime\over h})
&=(a-b)F_2(b^\prime;a-1,a^\prime;b,b^\prime;{k\over h},{k^\prime\over h})\cr&+(b-2a)F_2(b^\prime;a,a^\prime;b,b^\prime;{k\over h},{k^\prime\over h})\cr
&+aF_2(b^\prime;a+1,a^\prime;b,b^\prime;{k\over h},{k^\prime\over h}).}\eqno(3.19)$$
and finally we have
$$\eqalign{{b k\over h}F_2(b+1;a,a^\prime;b,b;{k\over h},{k^\prime\over h})
&=(a-b)(1-{k\over h})^{-a+1}(1-{k^\prime\over h})^{-a^\prime}{}_2F_1(a-1,a^\prime;b;{kk^\prime\over (h-k)(h-k^\prime)})\cr&+
(b-2a)(1-{k\over h})^{-a}(1-{k^\prime\over h})^{-a^\prime}{}_2F_1(a,a^\prime;b;{kk^\prime\over (h-k)(h-k^\prime)})\cr
&+a(1-{k\over h})^{-a-1}(1-{k^\prime\over h})^{-a^\prime}{}_2F_1(a+1,a^\prime;b;{kk^\prime\over (h-k)(h-k^\prime)}),}\eqno(3.20)$$
by means of (3.4). There are others recurrence relations for $F_2$ that can be develop in similar way, we mention three more:
$${k\over h}{(d-1)\over c} F_2(d;a,a^\prime;b+1,b^\prime+1;{k\over h},{k^\prime\over h})
=F_2(d-1;a,a^\prime;b,b^\prime+1;{k\over h},{k^\prime\over h})-F_2(d-1;a-1,a^\prime;b,b^\prime+1;{k\over h},{k^\prime\over h}),\eqno(3.21)$$

$${(a+1-b)} F_2(d;a,a^\prime;b,b^\prime;{k\over h},{k^\prime\over h})
=aF_2(d;a+1,a^\prime;b,b^\prime;{k\over h},{k^\prime\over h})-(b-1)F_2(d;a,a^\prime;b-1,b^\prime;{k\over h},{k^\prime\over h}).\eqno(3.22)$$

$$\eqalign{
F_2(d;a,a^\prime;b,b^\prime;{k\over h},{k^\prime\over h})
&={b(b-1)\over d-1}{k\over h}F_2(d-1;a,a^\prime;b,b^\prime;{k\over h},{k^\prime\over h})-{b(b-1)\over d-1}{k\over h}F_2(d-1;a,a^\prime;b-1,b^\prime;{k\over h},{k^\prime\over h})\cr
&+(b-a)F_2(d;a,a^\prime;b+1,b^\prime;{k\over h},{k^\prime\over h})
}\eqno(3.23)$$

%---------------------------------------------------------------------------
% 3. Applications
%---------------------------------------------------------------------------
\ni{\bf 4. Applications to perturbation theory}
\bigskip
%---------------------------------------------------------------------------
% 4.1. Singular Potentials
%---------------------------------------------------------------------------
\noindent Besides the important application of the integral (1.6) in computing the radial matrix elements of the electromagnetic interaction between the states of a relativistic electron in the Coulomb field, another important applications arises when we deal with singular Hamiltonians \sref{\hald-\halc}. First we consider the following lemma.
\medskip
%-----------------------
\noindent {\bf Lemma 10:}
%-----------------------
\item{1.} $F_2(d;-n,-m;b,b^\prime;x,1)= {(b^\prime-d)_m\over (b^\prime)_m} {}_3F_2(-n,d,1-b^\prime+d;b,1-b^\prime+d-m;x)$
\item{2.} $F_2(d;-n,-m;b,b^\prime;1,y)= {(b-d)_n\over (b)_n} {}_3F_2(-m,d,1-b+d;b^\prime,1-b+d-n;y)$
\item{3.} $F_2(d;-n,-m;b,b^\prime;1,1)= {(b^\prime-d)_m\over (b^\prime)_m} {}_3F_2(-n,d,1-b^\prime+d;b,1-b^\prime+d-m;1)$
\medskip
\noindent The proof of this lemma follows by means of the series representation (2.1) of the series $F_2$, therefor we omit it.
\medskip 
%--------------------------------
\ni{\bf 4.1. Singular Potentials}
%--------------------------------
\medskip
\noindent A family of quantum Hamiltonians known as generalized spiked harmonic oscillators
is given by the general Hamiltonian operator \sref{\hald-\halc}
$$
H=H_0+{\lambda\over x^\alpha},\quad H_0=-{d^2\over dx^2}+x^2+{A\over x^2}\quad (\lambda>0, \alpha>0,A\geq 0)\eqno(4.1)
$$
acting in the Hilbert space $L_2(0,\infty)$. Eigenfunctions $\psi\in L_2(0,\infty)$ of $H$ satisfy the Schr\"odinger equation
$$-\psi^{\prime\prime}+\bigg\{x^2+{A\over x^2}+{\lambda\over x^\alpha}\bigg\}\psi=E\psi\quad\hbox{ with } \psi(0)=0.$$
The function $\psi$ is an eigenfunction corresponding to the eigenvalue $E$ and the condition $\psi(0)=0$ is called a Dirichlet boundary condition. It is known that the unperturbed Hamiltonian $H_0$ admits exact solutions by means of the Gol'dman and Krivchenkov wavefunctions 
$$
\psi_n(x)=(-1)^n\sqrt{2(\gamma)_n\over n!\Gamma(\gamma)}
x^{\gamma-{1\over 2}}
e^{-{1\over 2}x^2}
{}_1F_1(-n;\gamma;x^2)\quad\hbox{ for }\quad n=0,1,2,\dots.\eqno(4.2)
$$
where $\gamma=1+{1\over 2}\sqrt{1+4A},$; and the exact eigenvalues are given by 
$$
E_n=2(2n+\gamma),\quad n =0,1,2,\dots\eqno(4.3)
$$
The orthonormality of the Gol'dman and Krivchenkov wavefunctions $
\{\psi_n(x)\}_{n=0}^\infty$ as well the computation of the matrix elements of the perturbed operator $V(x)=x^{-\alpha}$ in terms of $\{\psi_n(x)\}_{n=0}^\infty$ follows immediately by means of lemma 10. Indeed, we have
$$\int_0^\infty \psi_n(x)\psi_m(x)~dx = \delta_{mn},\quad n,m=0,1,2,\dots$$
as a result of 
$$\int\limits_0^\infty x^{2\gamma-1} e^{-h x^2}{}_1F_1(-n;\gamma;h x^2)
{}_1F_1(-m;\gamma;h x^2)~dx={1\over 2}{n!~\Gamma(\gamma)\over h^\gamma (\gamma)_n}\delta_{mn}.\eqno(4.4)
$$
The proof of the last integral reduces, after simple change of variables, to the computation of
$$\eqalign{F_2(\gamma;-n,-m;\gamma,\gamma;1,1)&=\sum\limits_{k=0}^m{(-m)_k\over k!}
{}_2F_1(-n,\gamma+k;\gamma;1)={1\over (\gamma)_n}
\sum\limits_{k=0}^m {(-m)_k(-k)_n\over  \ k!}={n!\over (\gamma)_n}\delta_{mn},}$$
since the product $(-m)_k(-k)_n$ differs from zero only for $n=k=m$.
Furthermore, the matrix elements of the singular operator $V(x)=x^{-\alpha}$  follow immediately by use of Lemma 10, and this yields\medskip
%% --------------------------
\noindent{\bf Lemma 11:}
%% --------------------------  
{\it If $2\gamma>\alpha$, then for all pairs of non-negative integers and $m$ and $n$ we have that 
$$\eqalign{\int\limits_0^\infty x^{2\gamma-\alpha-1} e^{-h x^2}{}_1F_1(-n,\gamma,h x^2)&
{}_1F_1(-m,\gamma,h x^2)dx=\cr
{h^{{\alpha\over 2}-\gamma}\over 2}{({\alpha\over 2})_n\Gamma(\gamma-{\alpha\over 2})\over (\gamma)_n}
&{}_3F_2(-m,{\gamma-{\alpha\over 2}},{1-{\alpha\over 2}};\gamma,1-{\alpha\over 2}-n;1).\cr}
\eqno(4.5)
$$}
Consequently, the matrix elements assume the explicit forms
$$<\psi_m|x^{-\alpha}|\psi_n>={(-1)^{n+m}\over (\gamma)_n}{\Gamma(\gamma-{\alpha\over 2})\over \Gamma(\gamma)}\sqrt{(\gamma)_n(\gamma)_m\over n!m!}{}_3F_2(-m,\gamma-{\alpha\over 2},1-{\alpha\over 2};\gamma,1-n-{\alpha\over 2};1)\eqno(4.6)$$
which can be used directly for a variational study of the perturbed Hamiltonian $H$. Note, in particular, for $n\neq 0$
$$<\psi_0|x^{-\alpha}|\psi_n>=(-1)^n\sqrt{(\gamma)_n\over n!}{\Gamma(\gamma-{\alpha\over 2})\over \Gamma(\gamma)}{({\alpha\over 2})_n\over (\gamma)_n}.$$
\bigskip
%---------------------------------------------------------------------------
% 4.2. Perturbed Kratzer Potential
%---------------------------------------------------------------------------
\ni{\bf 4.2. Perturbed Kratzer Potential}
\medskip
\noindent There has been great interest in the Coulomb potential with polynomial perturbation from both mathematical and physical points of view \sref{\fran}. However, the Kratzer potential \sref{\landf}
$$V(r)=-{B\over r}+{A\over r^2},\quad (B>0,A\geq 0)\eqno(4.7)$$
which contains the Coulomb potential as a particular case $(A=0)$ has apparently not been studied in such a context \sref{\fran}. The radial part of the Schr\"odinger equation for Kratzer potential reads
$$\bigg[-{d^2\over dr^2}-{2\over r}{d\over dr}+{l(l+1)\over r^2}+\bigg(-{B\over r}+{A\over r^2}\bigg)\bigg]\psi(r)=E\psi(r)\eqno(4.8)$$
The (un-normalized) wavefunctions are given explicitly by
$$\psi_{nl}(r)=r^se^{-{1\over 2}{B\over n+s+1}r}{}_1F_1(-n,2s+2,{B\over n+s+1}r),\quad n=0,1,2,\dots\eqno(4.9)$$
where
$$s=-{1\over 2}+{1\over 2}\sqrt{4A+(2l+1)^2}.\eqno(4.10)$$
The exact eigenvalues are given by the expression 
$$E_{nl}=-{B^2\over 4(n+s+1)^2},\quad n,l=0,1,2,\dots\eqno(4.11)$$
For the case of the Coulomb potential $A = 0$ and $s=l$, where $l=0,1,2,\dots$ is the angular momentum quantum number. The normalization constant (with the angular factor $4\pi$ omitted) 
appropriate for the wavefunction solutions (4.9) can be computed by means of the condition
$$\int_0^\infty |\psi_{nl}|^2r^2dr,$$
which leads to
$$C_{nl}^{-2}=\int_0^\infty r^{2s+2}e^{-{B\over n+s+1}r}[{}_1F_1(-n,2s+2,{B\over n+s+1}r)]^2dr$$
The integral just mentioned can be computed by an application of Lemma~10, which yields
$$C_{nl}^{-2}=2B^{-2s-3}(s+n+1)^{2s+4}{\Gamma(2s+2)\over (2s+2)_n}n!.\eqno(4.12)$$
If the potential (4.7) is perturbed by the operator $\lambda r^\alpha$, $\alpha>0$, that is to say if we consider
$$V(r)=-{B\over r}+{A\over r^2}+\lambda r^\alpha,\quad \lambda\geq 0,\alpha > 0$$
The matrix elements for the operator $r^{\alpha}$ by means of the (normalized) wavefunctions (4.9) and (4.12) are given through the integral
$$\eqalign{<\psi_{nl}|r^{\alpha}|\psi_{ml}>&=
C_{nl}C_{ml}\int_0^\infty r^{2s+2+\alpha}e^{-{B\over 2}({1\over n+s+1}+{1\over m+s+1})r}\times\cr
&{}_1F_1(-n;2s+2;{Br\over n+s+1}){}_1F_1(-m;2s+2;{Br\over m+s+1})dr}
$$
which yields by means of the series representation (2.1) of $F_2$ that   
$$\eqalign{&<\psi_{nl}|r^{\alpha}|\psi_{ml}>=\Gamma(2s+3+\alpha)~C_{nl}~C_{ml}~
\bigg[{B\over 2}\bigg({1\over n+s+1}+{1\over m+s+1}\bigg)\bigg]^{-\alpha-2s-3}
\cr
&\times\sum\limits_{k=0}^n{(2s+3+\alpha)_k(-n)_k\over (2s+2)_k~k!}\bigg({2m+2s+2\over 2s+(n+m)+2}\bigg)^k{}_2F_1(2s+3+\alpha+k,-m;2s+2;{2n+2s+2\over 2s+(n+m)+2}).}\eqno(4.13)
$$
In particular, the diagonal matrix elements (i.e. $m=n$) are given by
$$\eqalign{<\psi_{nl}|r^{\alpha}|\psi_{nl}>&={1\over 2}B^{-\alpha-2s}(n+s+1)^{\alpha+2s-1}{\Gamma(2s+3+\alpha)\over \Gamma(2s+2)}{(-1-\alpha)_n\over n!}\times\cr
&\quad {}_3F_2(-n,2s+3+\alpha,2+\alpha;2s+2,2+\alpha-n;1).
}\eqno(4.14)$$
The matrix elements (4.13) and (4.14) can also be used immediately, for example, for a variational study of the spectrum of $H = -\Delta + V(r),$ where $V(r)$ is a power $\sgn(q)r^{\alpha}.$
\bigskip
%---------------------------------------------------------------------------
% 5. Conclusion
%---------------------------------------------------------------------------
\ni{\bf 5. Conclusion:}
\bigskip

\noindent We have shown that a simple integral can be used to generate many identities that are very useful in mathematical and physical applications. It is clear that the confluent hypergeometric function  ${}_1F_1(-n,\alpha+1,x)$ can be expressed in terms of the associated Laguerre polynomials $L_n^\alpha(x)$ through the identity
$${}_1F_1(-n;b+1;x)={n!\over (b+1)_n}L_n^{b}(y)$$
and many of results dealing with the applications can be expressed equally well in terms of $L_n^\alpha(x)$. Some interesting new identities for class of intergrals containing pair of   associated Laguerre polynomials $L_n^\alpha(x)$ with weight measure  $\{t^{d-1}e^{-ht}\}$ 
can easily be derived on the basis of the explicit results presented here. 

\bigskip
\noindent {\bf Acknowledgments}
\medskip
\noindent Partial financial support of this work under Grant Nos. GP249507 and GP3438 from the 
Natural Sciences and Engineering Research Council of Canada is gratefully 
acknowledged.
\vfil\eject
%-------------------------------------
\references{1}
%-------------------------------------
\vfil\eject
%---------------------------------------------------------------------------
% Appendix I
%---------------------------------------------------------------------------
\ni{\bf Appendix I}
\bigskip
In this appendix we collect some results that follow by means of integral (2.4), and may be useful for further applications. Indeed, the possibilities implied by integral (2.4) are not limited to the present results; many other identities follow by use of the available expansions of the Gauss hypergeometric
functions ${}_2F_1$ in terms of elementary functions; see for instance \sref{\wfr}.

\noindent For $Re(d)>0$ and $|h|>|k|$
$$\int\limits_0^\infty t^{d-1}e^{-ht} {}_1F_1(a;d;k t)~ dt = {\Gamma(d)\over h^{d}}~{}_2F_1(d,a;d;{k\over h})= {\Gamma(d)\over h^{d}}~{}_1F_0(a;-;{k\over h})={\Gamma(d)\over h^{d}}~\bigg(1-{k\over h}\bigg)^{-a}.\eqno({\rm I}.1)$$
For $Re(d)>0$, $Re(h)>0$ and $a+d+1\neq 0,-2,-4,\dots$ 
$$\int\limits_0^\infty t^{d-1}e^{-ht} {}_1F_1(a;{a+d+1\over 2};{ht\over 2})~ dt 
={\Gamma(d)\over h^{d}}{}_2F_1(a,d;{a+d+1\over 2};{1\over 2})={\Gamma(d)\over h^{d}}~{\Gamma({1\over 2})\Gamma({a+d+1\over 2})\over \Gamma({1+d\over 2})\Gamma({1+a\over 2})}.\eqno({\rm I}.2)$$
For $Re(a)>0$, $Re(h)>0$ and $a+b\neq -2,-4,-6,\dots$ 
$$\eqalign
{\int\limits_0^\infty t^{a-1}e^{-ht} {}_1F_1(b;{a+b+2\over 2};{ht\over 2})~ dt 
&={\Gamma(a)\over h^{a}}{}_2F_1(a,b;{a+b+2\over 2};{1\over 2})\cr
&=2{\Gamma(a)\over h^{a}}~{\Gamma({1\over 2})\Gamma({a+b\over 2}+1)\over a-b}
\bigg[{1\over \Gamma({a\over 2})\Gamma({b+1\over 2})}-{1\over \Gamma({b\over 2})\Gamma({a+1\over 2})}\bigg].}\eqno({\rm I}.3)$$
For $Re(2a)>0$ and $|h|>|k|$,
$$\int\limits_0^\infty t^{2a-1}e^{-ht} {}_1F_1(a+1;a;k t)~ dt 
={\Gamma(2a)\over h^{2a}}~{}_2F_1(2a,a+1;a;{k\over k})=\Gamma(2a)~(h+k)(h-k)^{-2a-1}.\eqno({\rm I}.4)$$
For $Re(a)>-1$ and $|h|>|k|$,
$$\int\limits_0^\infty t^{a}e^{-ht} {}_1F_1(2a;a;k t)~dt 
={\Gamma(a+1)\over h^{a+1}}~\bigg(1+{k\over h}\bigg)\bigg(1-{k\over h}\bigg)^{-2a-1}.\eqno({\rm I}.5)$$
For $Re(a)>0$ and $|h|>|k|$,
$$\int\limits_0^\infty t^{a-1}e^{-ht} {}_1F_1(a+{1\over 2};2a;k t)~ dt 
={\Gamma(a)\over h^{a}}~{}_2F_1(a,a+{1\over 2};2a;{k\over h})={\Gamma(a)\over h^{a}}\bigg[{1-{k\over h}\bigg]^{-{1\over 2}}}~\bigg[{1\over 2}+{1\over 2}\sqrt{1-{k\over h}}\bigg]^{1-2a}.\eqno({\rm I}.6)$$
For $Re(a)>-{1\over 2}$ and $|h|>|k|$,
$$\int\limits_0^\infty t^{a-{1\over 2}}e^{-ht} {}_1F_1(a;2a;k t)~ dt 
={\Gamma(a+{1\over 2})\over h^{a+{1\over 2}}}\bigg[{1-{k\over h}\bigg]^{-{1\over 2}}}~\bigg[{1\over 2}+{1\over 2}\sqrt{1-{k\over h}}\bigg]^{1-2a}.\eqno({\rm I}.7)$$
For $Re(a)>0$ and $|h|>|k|$,
$$\int\limits_0^\infty t^{a-1}e^{-ht} {}_1F_1(a-{1\over 2};2a;k t)~ dt 
={\Gamma(a)\over h^{a}}~{}_2F_1(a,a-{1\over 2};2a;{k\over h})={\Gamma(a)\over h^{a}}~\bigg[{1\over 2}+{1\over 2}\sqrt{1-{k\over h}}\bigg]^{1-2a}.\eqno({\rm I}.8)$$
For $Re(a)>{1\over 2}$ and $|h|>|k|$,
$$\int\limits_0^\infty t^{a-{3\over 2}}e^{-ht} {}_1F_1(a;2a;k t)~ dt 
={\Gamma(a-{1\over 2})\over h^{a-{1\over 2}}}~\bigg[{1\over 2}+{1\over 2}\sqrt{1-{k\over h}}\bigg]^{1-2a}.\eqno({\rm I}.9)$$
For $Re(a)>0$ and $|h|>|k|$,
$$\int\limits_0^\infty t^{a-1}e^{-ht} {}_1F_1(a+{1\over 2};2a+1;k t)~ dt 
={\Gamma(a)\over h^{a}}~{}_2F_1(a,a+{1\over 2};2a+1;{k\over h})={\Gamma(a)\over h^{a}}~\bigg[{1\over 2}+{1\over 2}\sqrt{1-{k\over h}}\bigg]^{-2a}.\eqno({\rm I}.10)$$
For $Re(a)>-{1\over 2}$ and $|h|>|k|$,
$$\int\limits_0^\infty t^{a-{1\over 2}}e^{-ht} {}_1F_1(a;2a+1;k t)~ dt 
={\Gamma(a+{1\over 2})\over h^{a+{1\over 2}}}~\bigg[{1\over 2}+{1\over 2}\sqrt{1-{k\over h}}\bigg]^{-2a}.\eqno({\rm I}.11)$$
For $Re(a)>-1$ and $|h|>|k|$,
$$\eqalign{\int\limits_0^\infty t^{a}e^{-ht} {}_1F_1(a+{1\over 2};2a+1;k t)~ dt 
&={\Gamma(a+1)\over h^{a+1}}~{}_2F_1(a+1,a+{1\over 2};2a+1;{k\over h})\cr
&={\Gamma(a+1)\over h^{a+1}}\bigg[{1-{k\over h}\bigg]^{-{1\over 2}}}~\bigg[{1\over 2}+{1\over 2}\sqrt{1-{k\over h}}\bigg]^{-2a}.}\eqno({\rm I}.12)$$
For $Re(a)>-{1\over 2}$ and $|h|>|k|$,
$$\eqalign{\int\limits_0^\infty t^{a-{1\over 2}}e^{-ht} {}_1F_1(a+1;2a+1;k t)~ dt 
&={\Gamma(a+{1\over 2})\over h^{a+{1\over 2}}}\bigg[{1-{k\over h}\bigg]^{-{1\over 2}}}~\bigg[{1\over 2}+{1\over 2}\sqrt{1-{k\over h}}\bigg]^{-2a}.}\eqno({\rm I}.13)$$
For $Re(a)>-{1\over 2}$ and $|h|>|k|$,
$$\int\limits_0^\infty t^{a-{1\over 2}}e^{-ht} {}_1F_1(a;2a+1;k t)~ dt 
={\Gamma(a+{1\over 2})\over h^{a+{1\over 2}}}~\bigg[{1\over 2}+{1\over 2}\sqrt{1-{k\over h}}\bigg]^{-2a}.\eqno({\rm I}.14)$$
For $Re(a)>0$ and $|h|>|k|$,
$$\int\limits_0^\infty t^{a-1}e^{-ht} {}_1F_1(a+{1\over 2};2a+1;k t)~dt 
={\Gamma(a)\over h^a}~\bigg[{1\over 2}+{1\over 2}\sqrt{1-{k\over h}}\bigg]^{-2a}.\eqno({\rm I}.15)$$
For $|h|>|k|$ and $Re(a)>-1$
$$\int\limits_0^\infty t^{a}e^{-ht} {}_1F_1(1; 2;k t)~ dt 
={\Gamma(a+1)\over h^{a+1}}~{}_2F_1(a+1,1;2;{k\over h})={\Gamma(a)\over h}~\bigg[-{1\over h^{a}}+{1\over (h-k)^a}\bigg],\eqno({\rm I}.16)$$
and for $|h|>|k|$,
$$\int\limits_0^\infty e^{-ht} {}_1F_1(a+1; 2;k t)~ dt ={1\over h}~{}_2F_1(1,a+1;2;{k\over h})=\cases{-{1\over k}\log(1-{k\over h}),&if $a=0$\cr
\cr
{1\over k}\bigg[-{1\over a}+{h^a\over a(h-k)^a}\bigg],&if $a\neq 0$\cr}\eqno({\rm I}.17)
$$
In particular, for  $|h|>|k|$,
$$\int\limits_0^\infty e^{-ht} {}_1F_1(2;3;k t)~ dt = h^{-1}~{}_2F_1(1,2;3;{k\over h})=-{2\over k}\bigg[1+{h\over k}\ln(1-{k\over h})\bigg].\eqno({\rm I}.18)$$
For the particular calculations of the hypergeometric functions ${}_2F_1(a+1,1;2;x)$ in terms of elementary functions, we refer to Ref. \sref{\nas}. 

\noindent Direct aplications of (2.2) also yields:
For $Re(a)>0$ and $s>{k+k^\prime\over 2}$, we have
$$\int\limits_0^\infty t^{2a-1} e^{-st}{}_0F_1(-;a+{1\over 2};{k^2t^2\over 16}){}_0F_1(-;a+{1\over 2};{{k^\prime}^2t^2\over 16})dt={2^{2a}~\Gamma(2a)\over (4s^2-(k-k^\prime)^2)^{a}}{}_2F_1(a,a;2a;{4kk^\prime\over 4s^2-(k-k^\prime)^2})\eqno({\rm I}.19)$$
In particular
$$\int\limits_0^\infty t^{2a-1} e^{-st}\bigg[{}_0F_1(-;a+{1\over 2};{k^2t^2\over 16})\bigg]^2~dt={\Gamma(2a)\over s^{2a}}{}_2F_1(a,a;2a;{k^2\over s^2}).\eqno({\rm I}.20)$$
as a result of the identity
$${}_1F_1(a;2a;z)=e^{z\over 2}{}_0F_1(-;a+{1\over 2};{z^2\over 16}),$$
which also leads to the particular case
$$\int\limits_0^\infty t e^{-st}\bigg[{}_0F_1(-;{3\over 2};{k^2t^2\over 16})\bigg]^2~dt=-{1\over k^2}\log(1-{k^2\over s^2}).\eqno({\rm I}.21)$$

\end